# Metasurface array for single-shot spectroscopic ellipsometry

Shun Wen[1,#], Xinyuan Xue[1,#], Liqun Sun[1], Yuanmu Yang[1]*

[1]State Key Laboratory of Precision Measurement Technology and Instruments, Department of Precision Instrument, Tsinghua University, Beijing 100084, China

#These authors contributed equally.

*ymyang@tsinghua.edu.cn

**Spectroscopic ellipsometry is a potent method that is widely adopted for the measurement of thin film thickness and refractive index. However, a conventional ellipsometer, which utilizes a mechanically rotating polarizer and grating-based spectrometer for spectropolarimetric detection, is bulky, complex, and does not allow real-time measurements. Here, we demonstrated a compact metasurface array-based spectroscopic ellipsometry system that allows single-shot spectropolarimetric detection and accurate determination of thin film properties without any mechanical movement. The silicon-based metasurface array with a highly anisotropic and diverse spectral response is combined with iterative optimization to reconstruct the full Stokes polarization spectrum of the light reflected by the thin film with high fidelity. Subsequently, the film thickness and refractive index can be determined by fitting the measurement results to a proper material model with high accuracy. Our approach opens up a new pathway towards a compact and robust spectroscopic ellipsometry system for the high throughput measurement of thin film properties.**

For semiconductor processing, such as in the manufacturing of integrated circuits, flat display panels, and solar cells, to ensure processing quality and efficiency, it is crucial to accurately measure the thickness and optical constant of nanometer-scale thin films in a high throughput[1-3]. Instruments such as white-light interferometers, scanning electron microscopes, and atomic force microscopes may allow the measurement of film thickness with high accuracy, yet they have a relatively slow measurement speed and high system complexity. The measurement may also require direct contact with the sample under test. Spectroscopic ellipsometry is an alternative method that enables highly accurate and non-destructive measurements of thin film thickness as well as its optical constant, which has been widely implemented in semiconductor metrology and process monitoring[4-6].

In a typical spectroscopic ellipsometry system, it first measures the change in the polarization state of light reflected from the thin film under test. The wavelength-dependent complex reflectance ratio $\rho$ between $p$- and $s$-polarized light, or the ellipsometry parameter of the thin film is written as,

$$\rho = \tan(\Psi)\,e^{i\Delta}, \tag{1}$$

where $\Psi$ and $\Delta$ represent the amplitude ratio and phase difference, respectively. After experimentally obtaining the ellipsometry parameter, it is fitted to a theoretical model to eventually determine the thin film thickness and optical constant.

Nonetheless, a conventional spectroscopic ellipsometer typically modulates the polarization state via mechanical rotation of the compensator or analyzer, which has limited stability. Alternative polarization modulation approaches based on photoelastic or electro-optic effects may suffer from wavelength and temperature dependency. For spectral detection, it either requires wavelength scanning or the use of a multi-channel spectrometer[7]. As schematically shown in Fig. 1a, the resulting system becomes rather bulky and complex. Furthermore, the presence of rotating and/or scanning components leads to a slow measurement speed. Recently, alternative methods, such as dual-comb spectroscopic ellipsometry[8,9], have been proposed to partially address the abovementioned issues. However, dual-comb spectroscopic ellipsometry may have a limited spectral range for measurement, requires high-cost light sources, and also does not permit single-shot measurements.

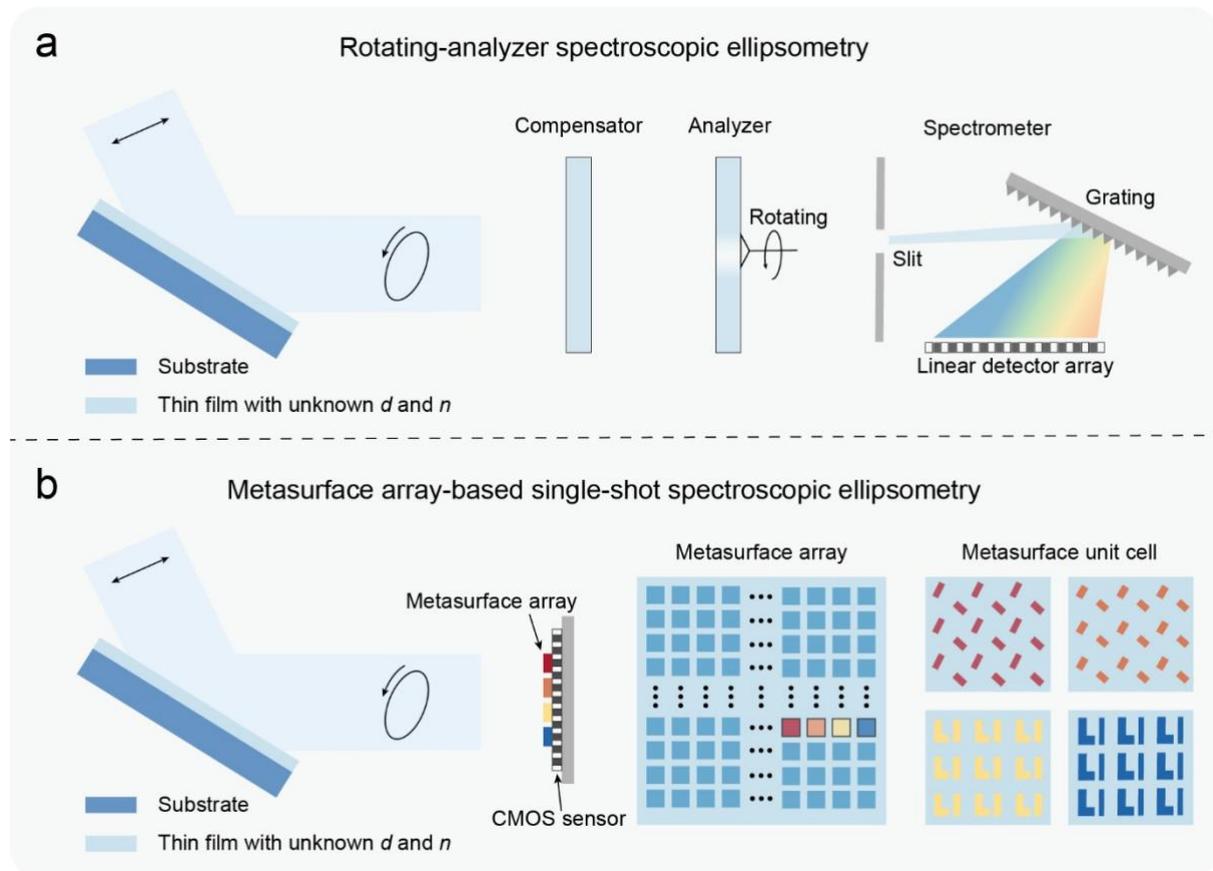

**Figure 1 | Comparison between conventional and metasurface array-based spectroscopic ellipsometry. a,** Schematics of a conventional spectroscopic ellipsometry system equipped with a compensator, a mechanically-rotating analyzer, and a grating-based multi-channel spectrometer. The thickness *d* and refractive index *n* of the thin film can be determined by fitting the measured ellipsometry parameters with a theoretical model. **b,** Schematics of a metasurface array-based single-shot spectroscopic ellipsometry system. The metasurface array, with each unit cell designed to support anisotropic and spectrally-diverse response, is used to encode the full Stokes polarization spectrum of light reflected from the thin film onto a CMOS image sensor. The ellipsometry parameters can be computationally reconstructed and used to determine the thin film properties. CMOS, complementary metal-oxide semiconductor.

Metasurface[10-13] is an emerging class of planar optical elements that allows extremely versatile manipulation of the amplitude[14,15], phase[16,17], polarization[18-20], and spectrum[21,22] of light at the subwavelength scale. Therefore, it may offer a new route toward constructing compact single-shot spectropolarimetric measurement systems. Recently, polarization-sensitive metalens arrays[23-26] and metasurface-based polarization gratings[27-29] have been utilized to build single-shot full Stokes polarization detection and imaging systems. On the other hand, to realize a compact spectrometer, one may design a metasurface-based narrowband filter array[30,31], with each filter responsible for transmitting a specific wavelength. Yet, a narrowband filter array intrinsically has a low light throughput. More recently, there has been a growing interest to develop metasurface-based miniaturized spectrometers[32-35] and hyperspectral imaging systems[36-39] based on computational reconstruction. In this approach, a metasurface filter array with a random spectral response is coupled with an iterative optimization algorithm or deep learning to reconstruct the spectrum, thus offering a higher light throughput and less stringent metasurface design requirement. For metasurface-based computational spectrometers, one of the prerequisites of the robust reconstruction of the spectral information is the construction of a filter array with a low spectral correlation and the accurate calibration of the filter response. However, due to the angle- and polarization-dependent response of most metasurface-based filters, accurate spectral reconstruction may become rather challenging for generic applications that involve light with a wide range of incident angles and polarization states.

Some recent studies have also aimed at using metasurface for simultaneous spectropolarimetric detection[40-45]. For instance, a metasurface-based polarization grating has been utilized for splitting light with different spectral and polarization components in the spatial domain[40,41]. However, similar to a conventional grating-based spectrometer, spectropolarimetry based on polarization grating has a fundamental tradeoff between the optical path length (system form factor) and the spectral resolution. We recently demonstrated computational spectropolarimetry based on a tunable liquid crystal-integrated metasurface[45], yet it requires active tuning elements and does not allow single-shot measurement. In addition, its spectral measurement range is limited to the near-infrared range.

In this work, we propose and experimentally demonstrate a single-shot spectroscopic ellipsometry system using a passive silicon-based metasurface array for spectropolarimetric encoding in the visible frequency regime, with the system schematically depicted in Fig.1b. The metasurface unit cell is meticulously designed to exhibit rich and anisotropic spectral features. Combining a single-shot measurement taken by the CMOS sensor with a straightforward iterative optimization algorithm, the full Stokes polarization spectrum of light reflected by the thin film can be reconstructed in high fidelity. Subsequently, the thickness and refractive index of the thin film can be obtained by fitting the measured ellipsometry parameters with a theoretical model. For the proposed ellipsometry system, light impinges on the metasurface filter array at a near-normal incident angle, resulting in a well-calibrated filter response, thus allowing the robust reconstruction of the spectropolarimetric information.

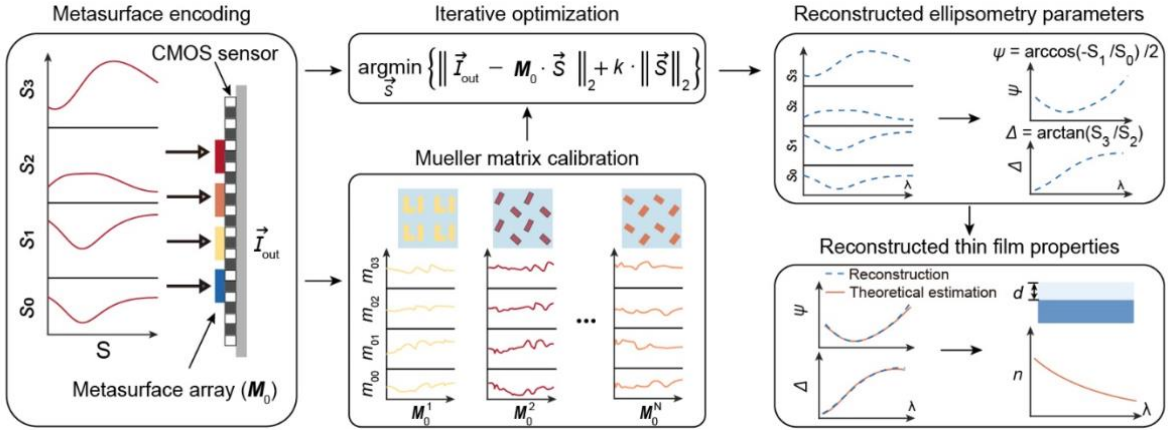

**Figure 2 | Working principle of the metasurface array-based single-shot spectroscopic ellipsometry.** The prototype consists of a single-layer metasurface array integrated on top of a CMOS image sensor. The Mueller matrix of the metasurface array ($M_0$) is pre-calibrated. Together with the measured intensity on the imaging sensor ($I_{out}$), the full Stokes polarization spectrum of the incident light can be reconstructed by a convex optimizer with $l_2$-regularization. In the following, the reconstructed full Stokes polarization spectrum

can be converted to the ellipsometry parameters for the determination of thin film thickness $d$ and refractive index $n$ by fitting the measurement with a multi-beam interference model.

The detailed working principle of the metasurface array-based spectroscopic ellipsometry is schematically illustrated in Fig. 2. The full Stokes polarization spectrum of light impinging on the metasurface array can be expressed as $\vec{S}(\lambda) = [s_0(\lambda), s_1(\lambda), s_2(\lambda), s_3(\lambda)]^T$. The metasurface array consists of $N$ elements and its polarization-dependent transmittance spectrum can be described by a Mueller matrix $M(\lambda)$, which is a function of the wavelength $\lambda$. The transmitted light with its intensity recorded by the CMOS sensor corresponds to the first element $s_0$ in a Stokes vector, which can be expressed as,

$$\vec{I}_{out} = \sum_{i=1}^{l} M_0(\lambda_i) \cdot \vec{S}(\lambda_i) = M_0 \cdot \vec{S} \qquad (2)$$

where $\vec{I}_{out}$ is an $N \times 1$ vector; $l$ is the number of spectral channels; $M_0 = [m_{00}, m_{01}, m_{02}, m_{03}]$ is the first row of $M(\lambda)$ (Supplementary Section 1). By minimizing the cost function with regularization, the full Stokes polarization spectrum can be reconstructed as,

$$\vec{S} = \underset{\vec{S}}{\operatorname{argmin}} \left\{ \left\| \vec{I}_{out} - M_0 \cdot \vec{S} \right\|_2 + k \cdot \left\| \vec{S} \right\|_2 \right\}. \qquad (3)$$

where $k$ is the regularization coefficient. Subsequently, the ellipsometry parameters can be calculated from $\vec{S}$ as (Supplementary Section 2),

$$\Psi = \frac{1}{2} \arccos\left(-\frac{s_1}{s_0}\right), \qquad (4)$$

$$\Delta = \arctan\left(\frac{s_3}{s_2}\right). \qquad (5)$$

After obtaining the ellipsometry parameters, the remaining steps for the determination of thin film properties are identical to that of a conventional ellipsometry system. We can first build a multi-beam interference model for the multilayer thin film stack under test and choose a proper material model to estimate the theoretical ellipsometry parameters (Supplementary Section 3). Thereafter, by minimizing the difference between the measurement and the theoretical estimation, one can determine the thickness $d$ and refractive index $n$ of the thin film.

In such a framework, the key to the accurate reconstruction of the ellipsometry parameters and thin film properties is to design a metasurface array with highly anisotropic and diverse spectral features, such that the correlation coefficient of each row of the Mueller matrix can be minimized. In this work, each element of the 20 × 20 metasurface array is made of 300-nm-thick silicon nanopillars on a sapphire substrate. The geometry of each element is optimized by minimizing the correlation coefficient of $M_0$ among different elements (Supplementary Section 4).

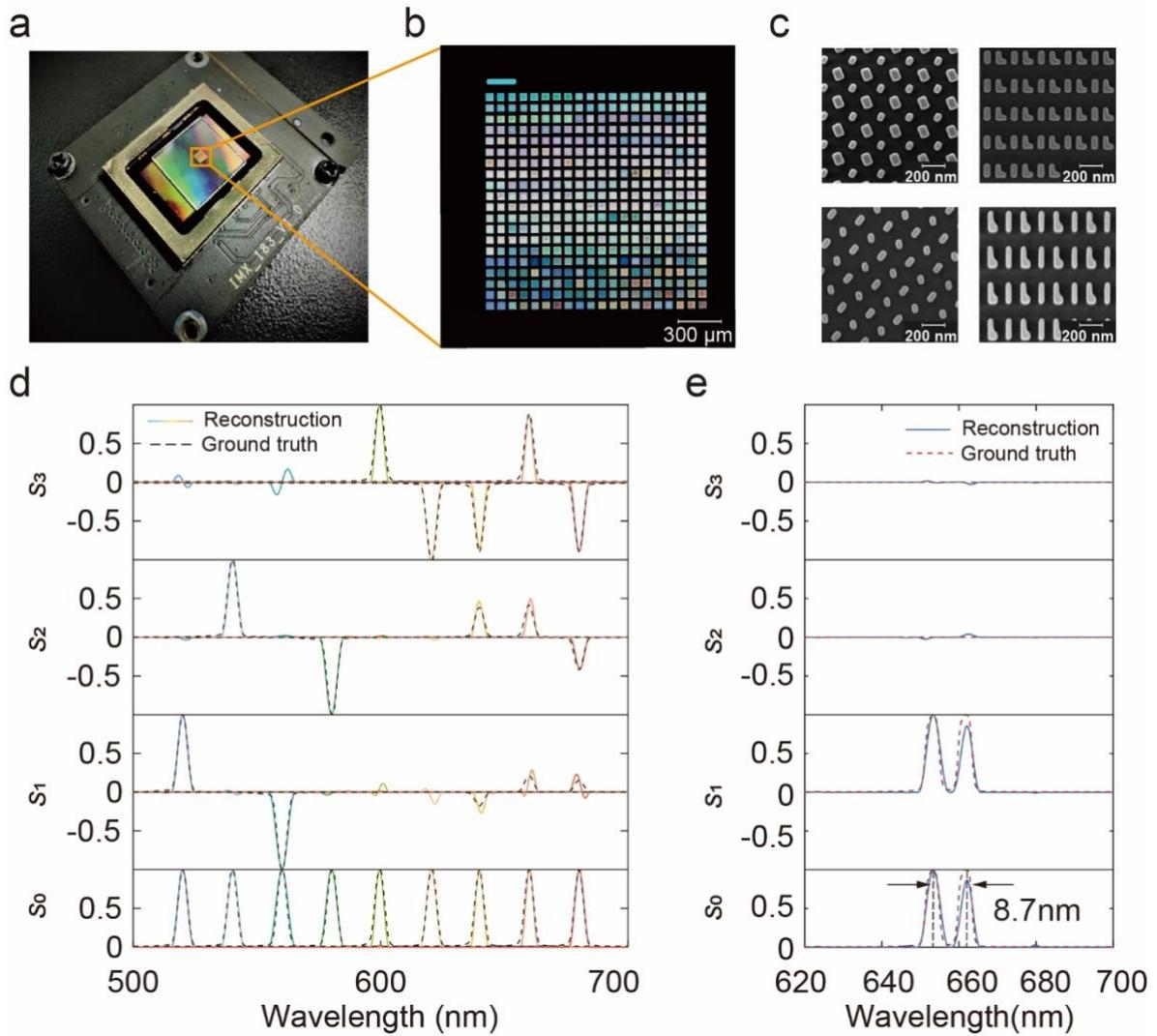

**Figure 3 | Characterization of the spectropolarimetric detection performance. a**, Photograph of the metasurface array-based spectroscopic ellipsometer. **b-c**, Zoom-in photograph (b) and SEM images (c) of the fabricated metasurface array. **d**, Full Stokes polarization spectra of narrowband light sources (with wavelength ranging from 520 nm to 680 nm in a 20-nm-interval, and with random polarization) measured with the metasurface array-based spectropolarimetric detection system (solid lines) compare with the ground truth measured with a quarter-waveplate, a rotating polarizer, and a conventional grating-based spectrometer (black dashed lines). **e**, Reconstructed full Stokes polarization spectrum of linearly polarized light with dual spectral peaks separated by 8.7 nm measured with the metasurface array-based spectropolarimetric detection system (blue solid line) compare with the ground truth measured with a quarter-waveplate, a rotating polarizer, and a conventional grating-based spectrometer (red dashed line). The peak wavelengths are highlighted by the black dashed line. SEM, scanning electron microscopy.

To experimentally demonstrate the metasurface array-based spectropolarimetric detection system, we assembled a prototype as shown in Fig. 3a. The metasurface array comprises 20 × 20 elements with a total size of 1.5 × 1.5 mm$^2$. The metasurface array was fabricated via the standard electron-beam lithography and reactive-ion etching process on a silicon-on-sapphire substrate (see Methods). The metasurface array was integrated onto a CMOS image sensor (Sony IMX-183) with a 5-μm-thick optically clear adhesive tape. The photograph of the metasurface array and 4 representative scanning electron microscopy (SEM) images of the metasurface elements are shown in Figs. 3b-c, respectively.

To calibrate $M_0$ of the metasurface array, we used a tunable monochromatic light source sampled at a 1-nm-interval across the spectral range from 500 nm to 700 nm (see Methods and Supplementary Fig. S6-S7). To evaluate the spectropolarimetric reconstruction performance of the system, we first reconstructed several spectra with simple narrowband features and with varying polarization states. As depicted in Fig. 3d, the reconstructed full Stokes polarization spectra agree very closely with the ground truth. The reconstruction error is quantitatively evaluated using the root mean square error (RMSE) = $\left\|\frac{\vec{S} - \vec{S'}}{4 \times v}\right\|_2$, where $\vec{S}$ is the normalized ground truth, $\vec{S'}$ is the normalized reconstruction result, and $v$ is the number of sampled spectral channels. The average peak-wavelength error, linewidth error, and RMSE, of the measured spectra are 0.17 nm, 0.33 nm, and 1.87%, respectively.

To further assess the system's capability to resolve fine spectral features, we prepared a linearly-polarized incident light with a double-peak spectrum by combining a semiconductor laser emitting at a wavelength of 653.5 nm with a white-light source coupled to a monochromator. As shown in Fig. 3e, the metasurface array-based spectropolarimetric detection system can clearly distinguish double spectral peaks at 653.5 nm and 662.2 nm separated by 8.7 nm. With a smaller spectral separation, the reconstructed spectrum gradually deviates from the ground truth (Supplementary Fig. S8). The relatively low spectral resolution of the system is due to the deterioration in the anisotropic spectral response of the experimentally realized metasurface array. This issue may be addressed via inverse design[46-48] of freeform metasurface unit cells with further decreased correlation and by improved metasurface fabrication process control. Despite the moderate spectral resolution of our system, it is worth noticing that the complexity of the full Stokes polarization spectrum of a thin film is a function of the film thickness and refractive index. For most thin films with a moderate thickness (< 1000 nm) and refractive index (< 3), the ellipsometry parameters can still be reconstructed with high fidelity.

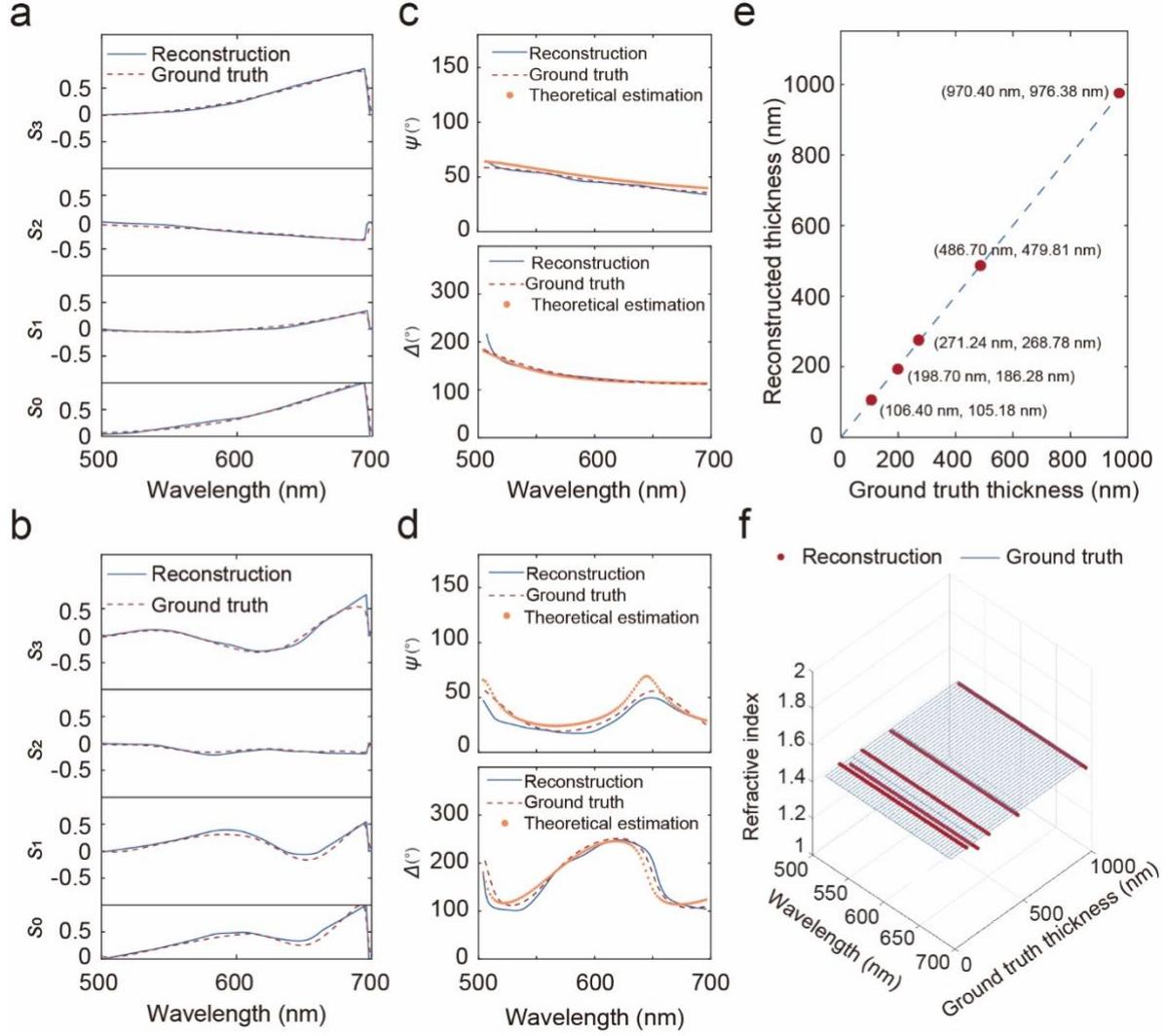

**Figure 4 | Measurement of the thickness and refractive index of SiO$_2$ thin films. a-b**, Comparison between the reconstructed full Stokes polarization spectra from the single-shot spectroscopic ellipsometer (blue solid line) and the ground truth (red dashed line) for SiO$_2$ thin film with a thickness of 100 nm (a) and 1000 nm (b), respectively. The ground truth is measured with a quarter-waveplate, a rotating polarizer, and a conventional grating-based spectrometer. **c-d**, Comparison among the ellipsometry parameters $\Psi$ and $\Delta$ from the single-shot spectroscopic ellipsometer (blue solid line), the ground truth (red dashed line), and the theoretical estimation (orange dots) for SiO$_2$ thin film with a thickness of 100 nm (c) and 1000 nm (d), respectively. The ground truth is measured with a quarter-waveplate, a rotating polarizer, and a conventional grating-based spectrometer. **e**, Comparison of the reconstructed thickness from the single-shot spectroscopic ellipsometer with the ground truth thickness from the commercial spectroscopic ellipsometer for five SiO$_2$ thin films under testing. **f**, Comparison of the reconstructed refractive index (red dots) from the single-shot spectroscopic ellipsometer with the ground truth (blue grids) from fused silica refractive index data in commercial spectroscopic ellipsometer for five SiO$_2$ thin films under testing.

Finally, to experimentally demonstrate spectroscopic ellipsometry measurement, five SiO$_2$ thin films with thicknesses ranging from 100 nm to 1000 nm deposited on a silicon substrate were selected as samples for testing, similar to the protocol in ref. 8. White light emitted from a halogen lamp, after passing through a 45° linear polarizer, impinged on the thin film samples at a 60° incident angle, close to the Brewster angle,
7

to ensure significant polarization conversion. The reflected light from the thin film impinged on the metasurface array at normal incidence (Supplementary Fig. S9). The reconstructed full Stokes polarization spectra of two representative $SiO_2$ thin films with thicknesses of 100 nm and 1000 nm are shown in Figs. 4a-b, respectively, which are compared to the ground truth measured with a quarter-waveplate, a rotating polarizer, and a conventional grating-based spectrometer, showing excellent agreement. Reconstructed spectra for thin films with other thicknesses can be found in Supplementary Fig. S10. A decent average RMSE of 4.5% was obtained for the five measured $SiO_2$ thin films.

Subsequently, the full Stokes polarization spectra can be converted to the ellipsometry parameters $\Psi$ and $\Delta$. The converted ellipsometry parameters of the $SiO_2$ thin film with thicknesses of 100 nm and 1000 nm are shown in Figs. 4c-d, respectively, which are compared with the ground truth and the theoretical estimation calculated from a three-layer model consisting of air, $SiO_2$, and a silicon substrate. Reconstructed ellipsometry parameters for films with other thicknesses can be found in Supplementary Fig. S10. By minimizing the squares error of the reconstruction and the theoretical estimation of $\Delta$, the thickness and refractive index of the $SiO_2$ thin films can be determined. The comparison between the film thickness reconstructed from our system and the measurement from a commercial spectroscopic ellipsometer (JA Woollam, V-VASE) is shown in Fig. 4e. Using the measured thickness from the commercial spectroscopic ellipsometer as the ground truth, the accuracy of the thickness measurement, defined as the relative error between the reconstructed film thickness and the ground truth, is only 2.16% on average for the five $SiO_2$ thin films. The precision of the thickness measurement, defined as the standard deviation in the 10 measurements within a 10-minute timeframe, is only 1.28 nm on average for the five $SiO_2$ thin films (Supplementary Fig. S11). In the fitting process, the refractive index of the $SiO_2$ thin films was assumed to follow the Cauchy model as $n = A + \frac{B\lambda^2}{\lambda^2 - C^2} - D\lambda^2$, where $A$, $B$, $C$ and $D$ are the fitting parameters[49]. The reconstructed refractive index dispersions for the five $SiO_2$ thin films are shown in Fig. 4f, with an average accuracy and precision of 0.84% and 0.0032, respectively (Supplementary Fig. S11). The entire computational reconstruction process was implemented in MATLAB, taking ~800 ms to reconstruct the full Stokes polarization spectrum and ~200 ms to reconstruct the thin film thickness and refractive index on a desktop computer equipped with an AMD Ryzen 7 3700x CPU and 32 GB RAM. To further enhance the speed and accuracy of the reconstruction, one may implement deep learning-based algorithms for both steps[34,36,50,51].

In conclusion, we have proposed and experimentally demonstrated a new type of metasurface array-based spectroscopic ellipsometry system for the single-shot measurement of thin film properties. Compared to conventional spectroscopic ellipsometers, our system has no mechanical moving parts or phase-modulating elements, Consequently, it may enable high-throughput, online measurement of thin film properties in semiconductor processing, such as in a thin film etching or deposition system. In the current prototype, we only reconstructed the thickness and refractive index of a single-layer lossless film. To allow the reconstruction of multi-layer films with losses, the design and fabrication of the metasurface array, as well as the ellipsometry fitting model, shall be further improved. With a modified metasurface design, the operation of the metasurface array-based spectroscopic ellipsometry system may be extended to a wide spectral range spanning from the ultraviolet to the terahertz band. The metasurface array also holds promise for spectropolarimetric imaging, which may further allow the non-destructive characterization of spatially-inhomogeneous thin films.

## Methods

**Metasurface fabrication**. The fabrication of the metasurface was done via a commercial service offered by Tianjin H-Chip Technology Group. The fabrication process is schematically shown in Supplementary Fig. S4. The metasurface array was first patterned on a 300-nm-thick monocrystalline silicon film on a sapphire substrate by the electron beam lithography using a negative tone resist hydrogen silsesquioxane (HSQ). The next step involved transferring the pattern onto the silicon layer using HSQ as the mask via the reactive ion etching process. Finally, the HSQ resist is removed by buffered oxide etchant.

**Metasurface characterization**. The experiment setup for the calibration of the metasurface array response is schematically shown in Supplementary Fig. S5. A supercontinuum laser (YSL Photonics, SC-Pro-7) was coupled to a monochromator (Zolix, Omni-λ2007i) and a collimation lens to emit collimated light with a narrowband spectrum. To eliminate spatial coherence and the resulting speckles on the imaging sensor, the light passed through a rotating diffuser before reaching the metasurface array. To ensure uniform illumination on the metasurface array, we established a Köhler illumination condition using two lenses with focal lengths of 50 mm and 35 mm, respectively, along with two apertures.

To calibrate $M_0$ of the metasurface array, its transmittance was measured with 0° linear polarization ($T_x$), 45° linear polarization ($T_{45}$), 90° linear polarization ($T_y$), and left-handed circular polarization ($T_{lcp}$), respectively. The polarization state of the incident light was adjusted by a polarizer and a broadband quarter-waveplate. For each sampled spectral channel, the grayscale image of the metasurface array was captured by the CMOS sensor. The transmittance was subsequently calculated by dividing the light intensity underneath the metasurface array by the light intensity in a blank region after the subtraction of the background noise. $M_0$ of the metasurface array was then calculated as,

$$M_0 = \begin{bmatrix} \frac{1}{2}(T_x + T_y) \\ \frac{1}{2}(T_x - T_y) \\ T_{45} - \frac{1}{2}(T_x + T_y) \\ \frac{1}{2}(T_x + T_y) - T_{lcp} \end{bmatrix}^T . \tag{6}$$

## Data availability
All relevant data are available in the main text, in the Supporting Information, or from the authors.

## Acknowledgment
This work was supported by the National Natural Science Foundation of China (62135008, 61975251) and by the Guoqiang Institute, Tsinghua University.

## Author contributions
Y.Y., S.W. and X.X. conceived this work. S.W. and X.X. designed and characterized the metasurface, developed the spectropolarimetric reconstruction algorithm, built the metasurface array-based spectroscopic ellipsometer system, and measured the thin film properties; S.W., X.X., L.S, and Y.Y. analyzed the data; S.W., X.X., and Y.Y. wrote the manuscript. Y.Y. supervised the project.

## Competing interests
The authors declare no competing interests.

# Supplementary Information:

# Metasurface array for single-shot spectroscopic ellipsometry

Shun Wen[1,#], Xinyuan Xue[1,#], Liqun Sun[1], Yuanmu Yang[1]*

[1]State Key Laboratory of Precision Measurement Technology and Instruments, Department of Precision Instrument, Tsinghua University, Beijing 100084, China

#These authors contributed equally.

*ymyang@tsinghua.edu.cn

## 1. Measurement of the full Stokes polarization spectrum

For polarized light impinging on the metasurface array, it can be expressed in the form of the full Stokes polarization spectrum as $\vec{S}(\lambda) = [s_0(\lambda), s_1(\lambda), s_2(\lambda), s_3(\lambda)]^T$. The polarization-dependent transmittance spectra of the metasurface elements can be described by a 4 × 4 Mueller matrix ($M$). The full Stokes polarization spectrum of the transmitted light can be written as,

$$\vec{S}_{out} = \begin{bmatrix} \vec{s}_{out0} \\ \vec{s}_{out1} \\ \vec{s}_{out2} \\ \vec{s}_{out3} \end{bmatrix} = \sum_{i=1}^{l} M(\lambda_i)\vec{S}(\lambda_i) = \begin{bmatrix} m_{00}(\lambda_i) & m_{01}(\lambda_i) & m_{02}(\lambda_i) & m_{03}(\lambda_i) \\ m_{10}(\lambda_i) & m_{11}(\lambda_i) & m_{12}(\lambda_i) & m_{13}(\lambda_i) \\ m_{20}(\lambda_i) & m_{21}(\lambda_i) & m_{22}(\lambda_i) & m_{23}(\lambda_i) \\ m_{30}(\lambda_i) & m_{31}(\lambda_i) & m_{32}(\lambda_i) & m_{33}(\lambda_i) \end{bmatrix} \begin{bmatrix} s_0(\lambda_i) \\ s_1(\lambda_i) \\ s_2(\lambda_i) \\ s_3(\lambda_i) \end{bmatrix}, \quad (S1)$$

where $l$ is the number of spectral channels;

Since the CMOS sensor can only record the light intensity, which corresponds to the element $\vec{s}_{out0}$ in $\vec{S}_{out}$, the light intensity detected by the CMOS sensor can be represented by the first element $\vec{s}_{out0}$ of the Stokes vector as,

$$\vec{I}_{out} = \vec{s}_{out0} = \sum_{i=1}^{l} M_0(\lambda_i)\vec{S}(\lambda_i), \quad (S2)$$

where $M_0(\lambda_i) = [m_{00}(\lambda_i), m_{01}(\lambda_i), m_{02}(\lambda_i), m_{03}(\lambda_i)]$ is the first row of $M(\lambda_i)$; $m_{00}$, $m_{01}$, $m_{02}$, and $m_{03}$ are all $N \times l$ matrices, and $N$ is the number of metasurface elements.

## 2. Derivation of ellipsometry parameters from the measured Stokes vector

Light reflected from a thin film can be represented by the Jones vector, as schematically shown in Fig. S1a, as,

$$\begin{bmatrix} E'_p \\ E'_s \end{bmatrix} = \begin{bmatrix} R_p E_p \\ R_s E_s \end{bmatrix} = \begin{bmatrix} R_p|E_p|e^{i\phi} \\ R_s|E_s| \end{bmatrix}, \quad (S3)$$

where $\begin{bmatrix} E_p \\ E_s \end{bmatrix}$, $\begin{bmatrix} E'_p \\ E'_s \end{bmatrix}$ is the Jones vector of the incident and reflected light, respectively; $R_p$ and $R_s$ are the complex reflection coefficients of the thin film for $p$- and $s$-polarized light, respectively; and $\phi$ is the phase difference between the $p$- and $s$-polarized light.

The Stokes vector of light reflected by the thin film can be derived from Eq. (S3) as,

$$\begin{bmatrix} s_0 \\ s_1 \\ s_2 \\ s_3 \end{bmatrix} = \begin{bmatrix} |R_p E_p|^2 + |R_s E_s|^2 \\ |R_p E_p|^2 - |R_s E_s|^2 \\ 2R_p R_s |E_p||E_s|\cos(\phi) \\ 2R_p R_s |E_p||E_s|\sin(\phi) \end{bmatrix} = \begin{bmatrix} |R_p E_p|^2 + |R_s E_s|^2 \\ |R_p E_p|^2 - |R_s E_s|^2 \\ 2|R_p||R_s||E_p||E_s|\cos(\phi + \Delta) \\ 2|R_p||R_s||E_p||E_s|\sin(\phi + \Delta) \end{bmatrix},\quad (S4)$$

where $\Delta$ is the phase difference of $R_p$ and $R_s$.

We set the incident light to linearly polarized at 45° in the experiment. Therefore, $\phi = 0$ and $E_p = E_s$. Combining Eq. (S4) with Eq. (1) in the main text, one can derive that,

$$\frac{s_0}{s_1} = \frac{|R_p|^2 + |R_s|^2}{|R_p|^2 - |R_s|^2} = \frac{[\tan(\Psi)]^2 + 1}{[\tan(\Psi)]^2 - 1} = -\frac{1}{\cos(2\Psi)}, \quad (S5)$$

$$\frac{s_2}{s_3} = \frac{\cos(\Delta)}{\sin(\Delta)} = \frac{1}{\tan(\Delta)}. \quad (S6)$$

Consequently, the ellipsometry parameters can be written as,

$$\Psi = \frac{1}{2}\arccos\left(-\frac{s_1}{s_0}\right), \quad (S7)$$

$$\Delta = \arctan\left(\frac{s_3}{s_2}\right). \quad (S8)$$

## 3. Derivation of the theoretical ellipsometry parameter

According to the Fresnel equations, $R_p$ and $R_s$ can be expressed as,

$$R_p = \frac{r_{01}^p + r_{12}^p e^{-i2\delta_p}}{1 + r_{12}^p \cdot r_{01}^p e^{-i2\delta_p}}, \quad R_s = \frac{r_{01}^s + r_{12}^s e^{-i2\delta_s}}{1 + r_{12}^s \cdot r_{01}^s e^{-i2\delta_s}}, \quad (S9)$$

where $r_{01}$ and $r_{12}$ is the reflectance ratio of the interface; $\delta_p$ and $\delta_s$ are the phase change of reflected $p$- and $s$-polarized light, respectively. For a single-layer thin film, as shown in Fig. S1b, $r_{01}$, $r_{12}$, $\delta_p$ and $\delta_s$ can be expressed as,

$$r_{01}^p = \frac{n_1 \cos\theta_0 - n_0 \cos\theta_1}{n_1 \cos\theta_0 + n_0 \cos\theta_1}, \quad r_{12}^p = \frac{n_2 \cos\theta_1 - n_1 \cos\theta_2}{n_2 \cos\theta_1 + n_1 \cos\theta_2}, \quad (S10)$$

$$r_{01}^s = \frac{n_0 \cos\theta_0 - n_1 \cos\theta_1}{n_0 \cos\theta_0 + n_1 \cos\theta_1}, \quad r_{12}^s = \frac{n_1 \cos\theta_1 - n_2 \cos\theta_2}{n_1 \cos\theta_1 + n_2 \cos\theta_2}, \quad (S11)$$

$$\delta_p = \delta_s = 2\pi\left(\frac{d}{\lambda}\right) n_1 \cos\left\{\sin^{-1}\left[\frac{n_2}{n_1}\sin(\theta_2)\right]\right\}. \quad (S12)$$

According to Eqs. (S9)-(S12), $R_p$ and $R_s$ are related to the film thickness $d$ and refractive index $n$. Therefore, the thickness and refractive index can be obtained by calculating the reflectance ratio as,

$$\rho = \frac{R_p}{R_s} = \tan(\Psi)e^{i\Delta}, \quad (S13)$$

where $\Psi$ and $\Delta$ are the ellipsometry parameters.

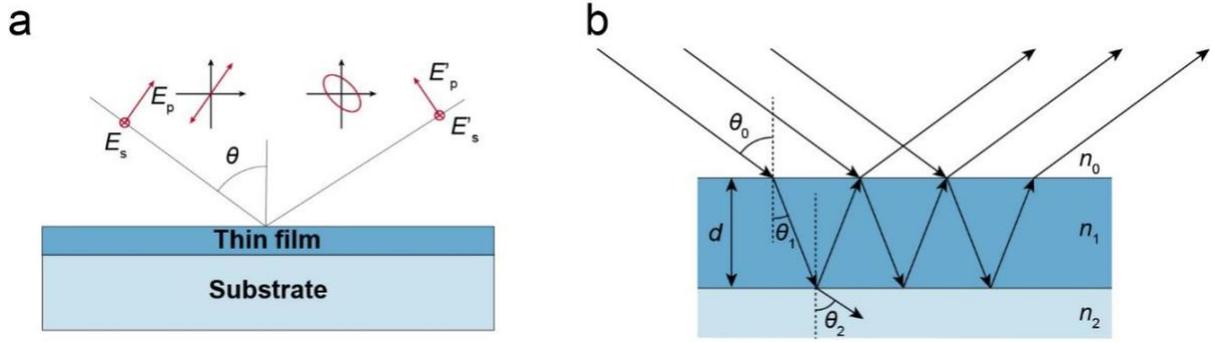

**Figure S1 | a,** The linearly polarized incident light is converted to an unknown (typically elliptic) polarization state upon reflection from the thin film under test. **b,** Schematic diagram of the multi-beam interference model for a single-layer thin film.

## 4. Method for designing metasurface elements

To obtain metasurface elements with anisotropic and diverse spectral features, we used Lumbrical FDTD to establish a database with more than 6000 elements. All elements were designed based on two types of di-atomic nanopillar patterns with broken in-plane symmetry, as depicted in Fig. S2. By changing the periodicity of the unit cell and the geometric parameters of the nanopillars, we can obtain metasurface elements with distinctive anisotropic spectral responses.

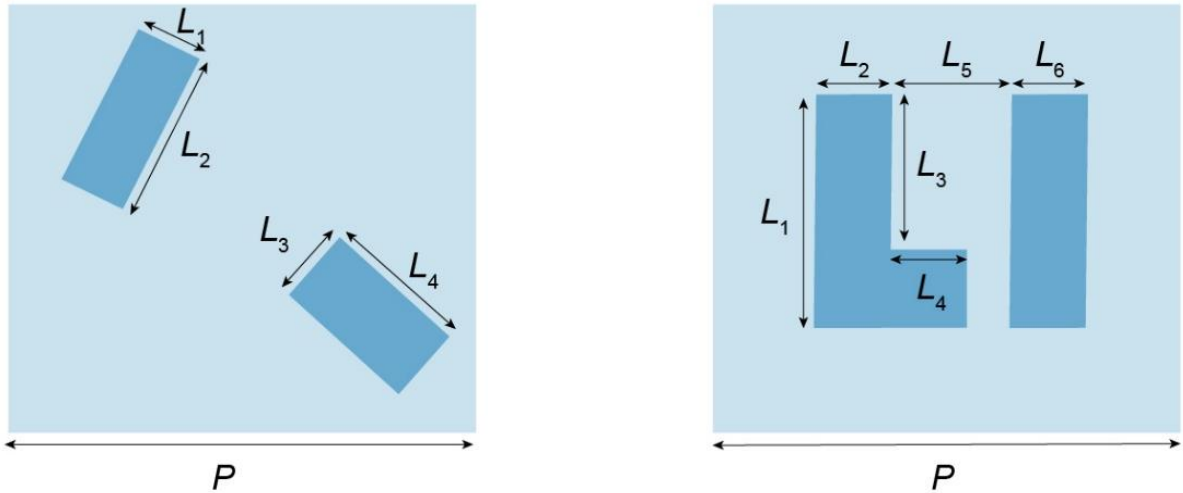

**Figure S2 |** Schematic diagram of two types of metasurface unit cells and the definition of parameters.

To ensure the fidelity of full Stokes polarization spectrum reconstruction, the correlation coefficients of each row of the Mueller matrix should be minimized. The correlation coefficient is defined as,

$$c_{i,j} = \frac{\mathrm{cov}\left(M_0^i(\lambda), M_0^j(\lambda)\right)}{\sigma_{M_0^i(\lambda)} \sigma_{M_0^j(\lambda)}}, \qquad (S14)$$

where $c_{i,j}$ is the correlation coefficient between $M_0^i(\lambda)$ and $M_0^j(\lambda)$; $i$ and $j$ represent transmittance responses of the *i*-th and *j*-th metasurface elements, respectively. cov and $\sigma$ represent the covariance operation and the standard deviation, respectively.

To design an array of 20 × 20 elements, two elements with the lowest correlation coefficient are first selected as the starting point. Subsequently, the correlation coefficients between all the remaining elements and the initial two elements are calculated. The element with the smallest average correlation coefficient is chosen as the third element. This process is repeated iteratively until all 20 × 20 elements are obtained. The correlation coefficients of the finally obtained $M_0(\lambda)$ is $C_0 = [C_{00}, C_{01}, C_{02}, C_{03}]$, which correspond to the average correlation coefficient of $[m_{00}, m_{01}, m_{02}, m_{03}]$, as shown in Fig. S3.

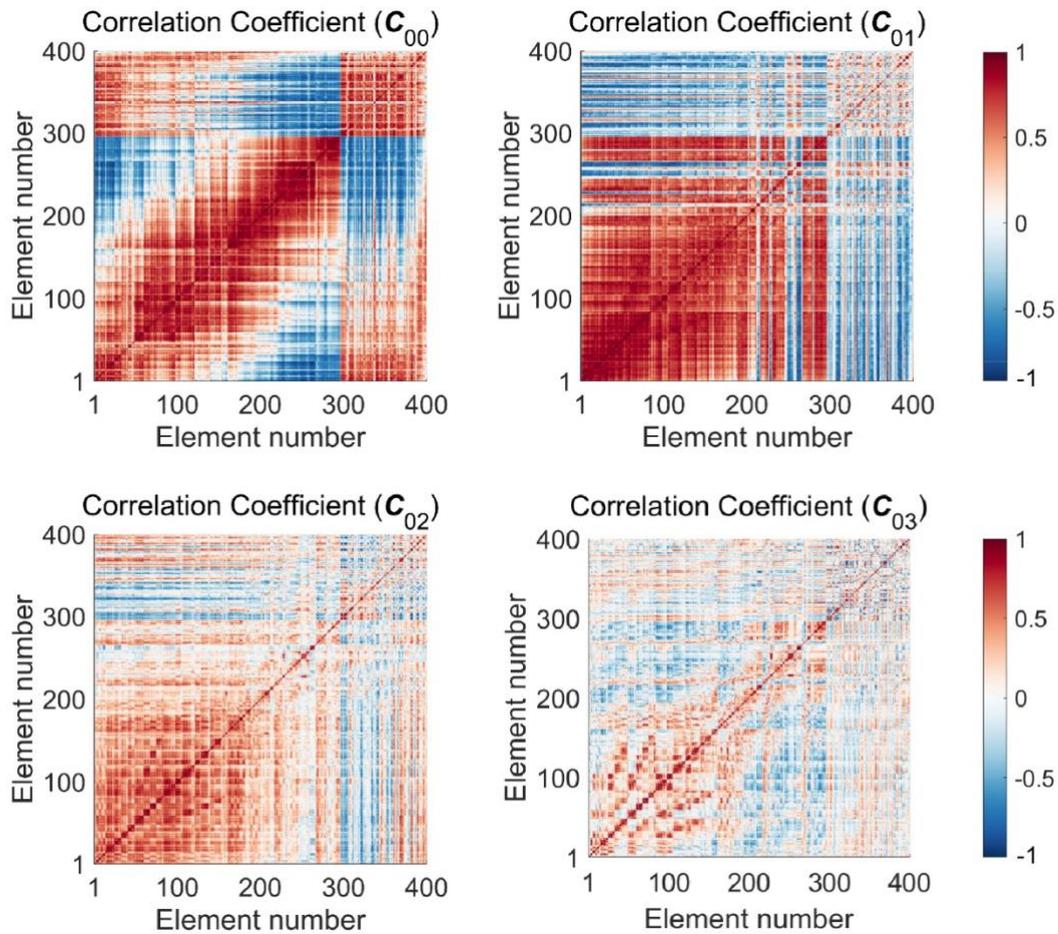

**Figure S3** | Calculated correlation coefficients $C_{00}$, $C_{01}$, $C_{02}$ and $C_{03}$ of the first row of the Mueller matrix over different metasurface elements.

## 5. Metasurface fabrication process

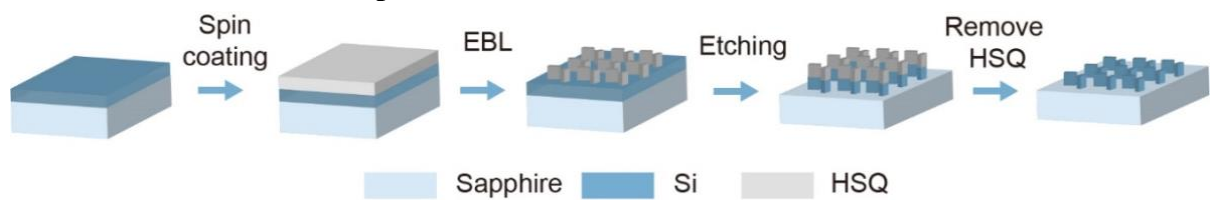

**Figure S4 |** Schematic of the fabrication process flow chart of the metasurface. EBL, Electron beam lithography; HSQ, Hydrogen silsesquioxane.

## 6. Experimental setup for calibration

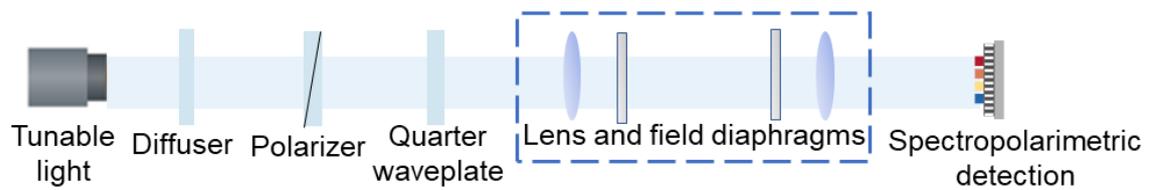

**Figure S5 |** Experimental setup for the calibration of $M_0$ of the metasurface array.

## 7. The experimentally calibrated $M_0$ of the metasurface array.

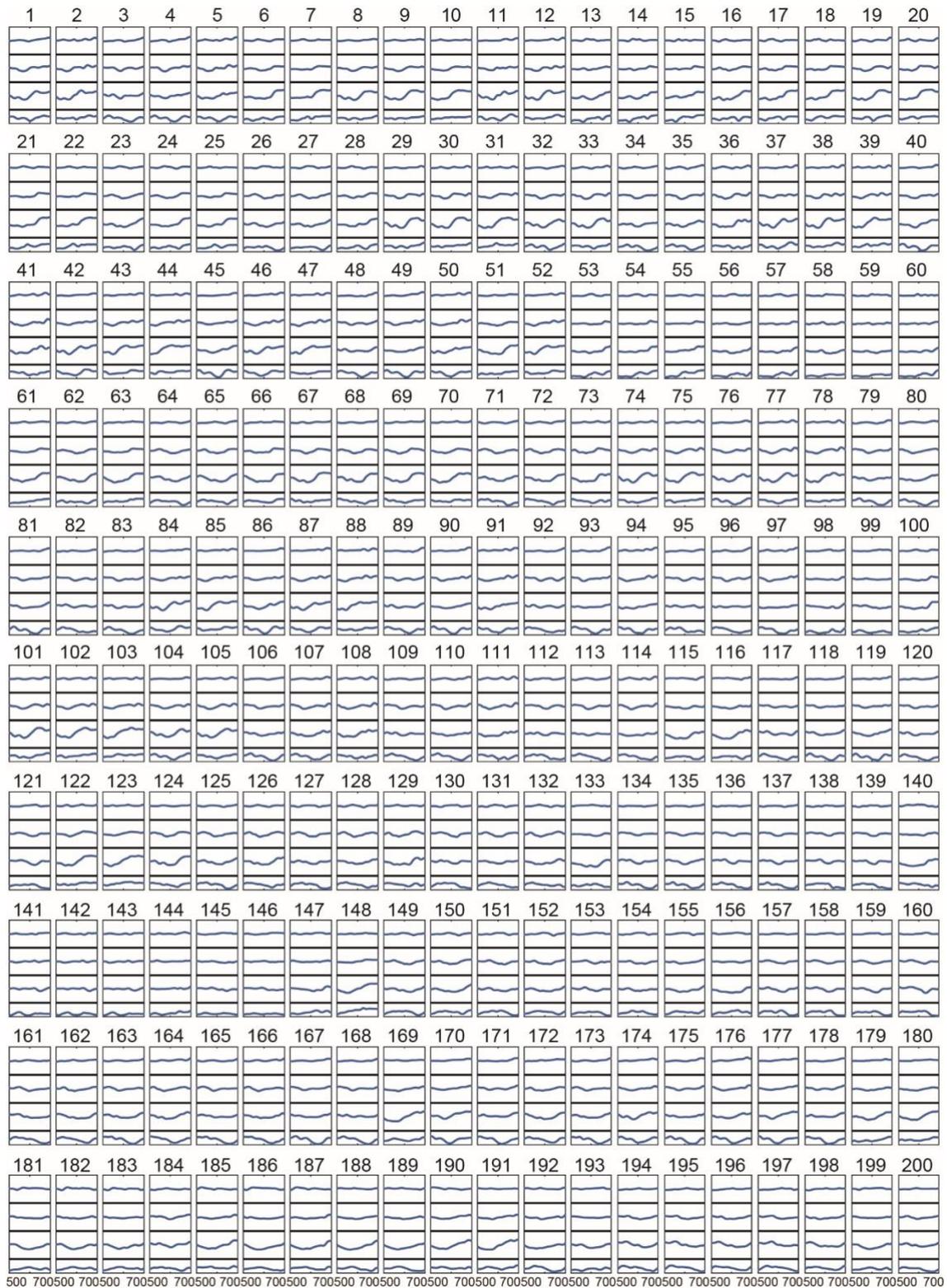

**Figure S6 | The experimentally calibrated $M_0^i$ of 1st – 200th metasurface elements.** In each subplot, the horizontal axis represents wavelength (in nanometers), and the vertical axis represents $m_{00}$, $m_{01}$, $m_{02}$, and $m_{03}$ from bottom to top, respectively.

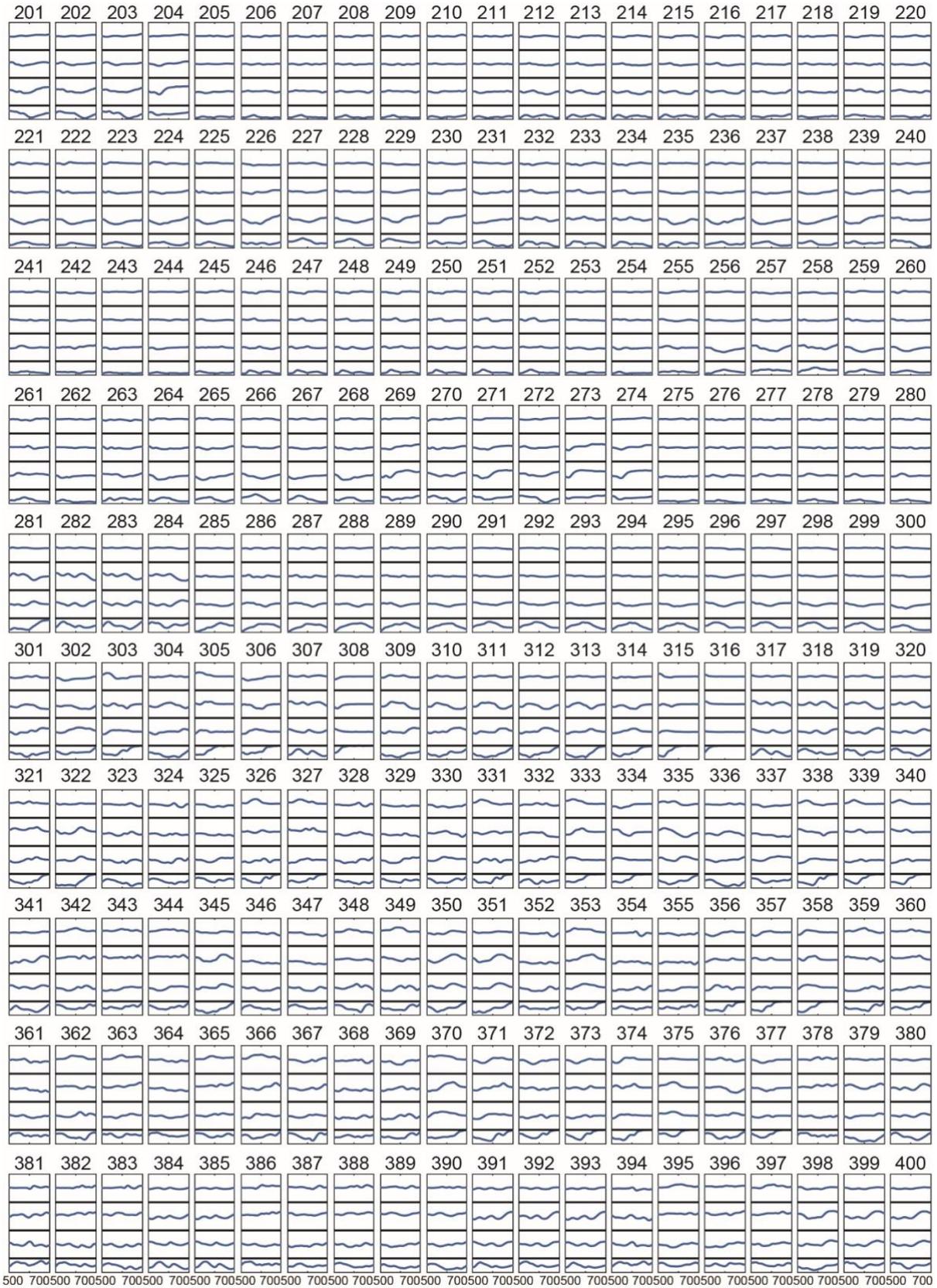

**Figure S7 | The calibrated $M_0^i$ of 201st – 400th metasurface elements.** In each subplot, the horizontal axis represents wavelength (in nanometers), and the vertical axis represents $m_{00}$, $m_{01}$, $m_{02}$, and $m_{03}$ from bottom to top, respectively.

## 8. Spectropolarimetric reconstruction of dual-peak spectra

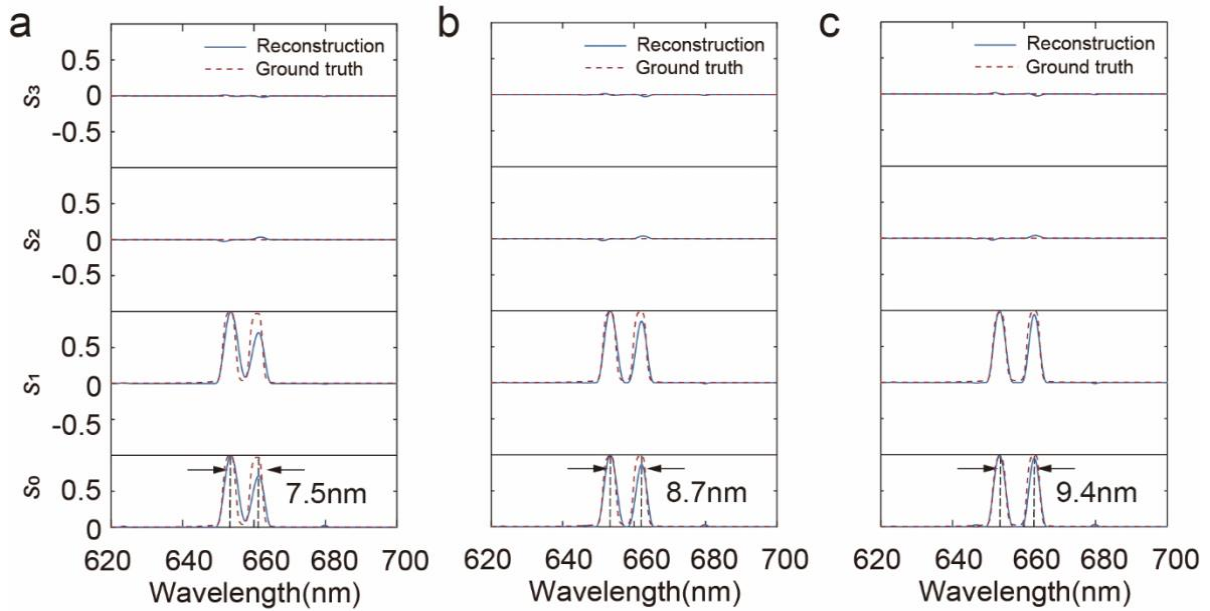

**Figure S8 | Characterization of the spectral resolution of the spectropolarimetric detection system. a-c**, Reconstructed full Stokes polarization spectrum of linearly polarized light with dual spectral peaks separated by 7.5 nm (a), 8.7 nm (b), and 9.4 nm (c), respectively, measured with the metasurface array-based spectropolarimetric detection system (blue solid line) compare with the ground truth measured with a quarter-waveplate, a rotating polarizer, and a conventional grating-based spectrometer (red dashed line). The peak wavelengths are highlighted by the black dashed line.

## 9. Experimental setup for measuring the thin film properties

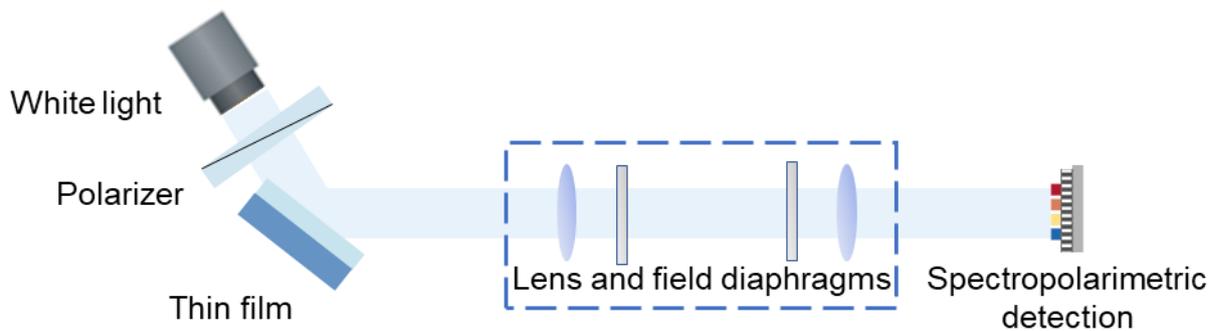

**Figure S9 |** Experimental setup for measuring the thin film properties.

# 10. The reconstructed full Stokes polarization spectra and ellipsometry parameters of the SiO$_2$ thin films.

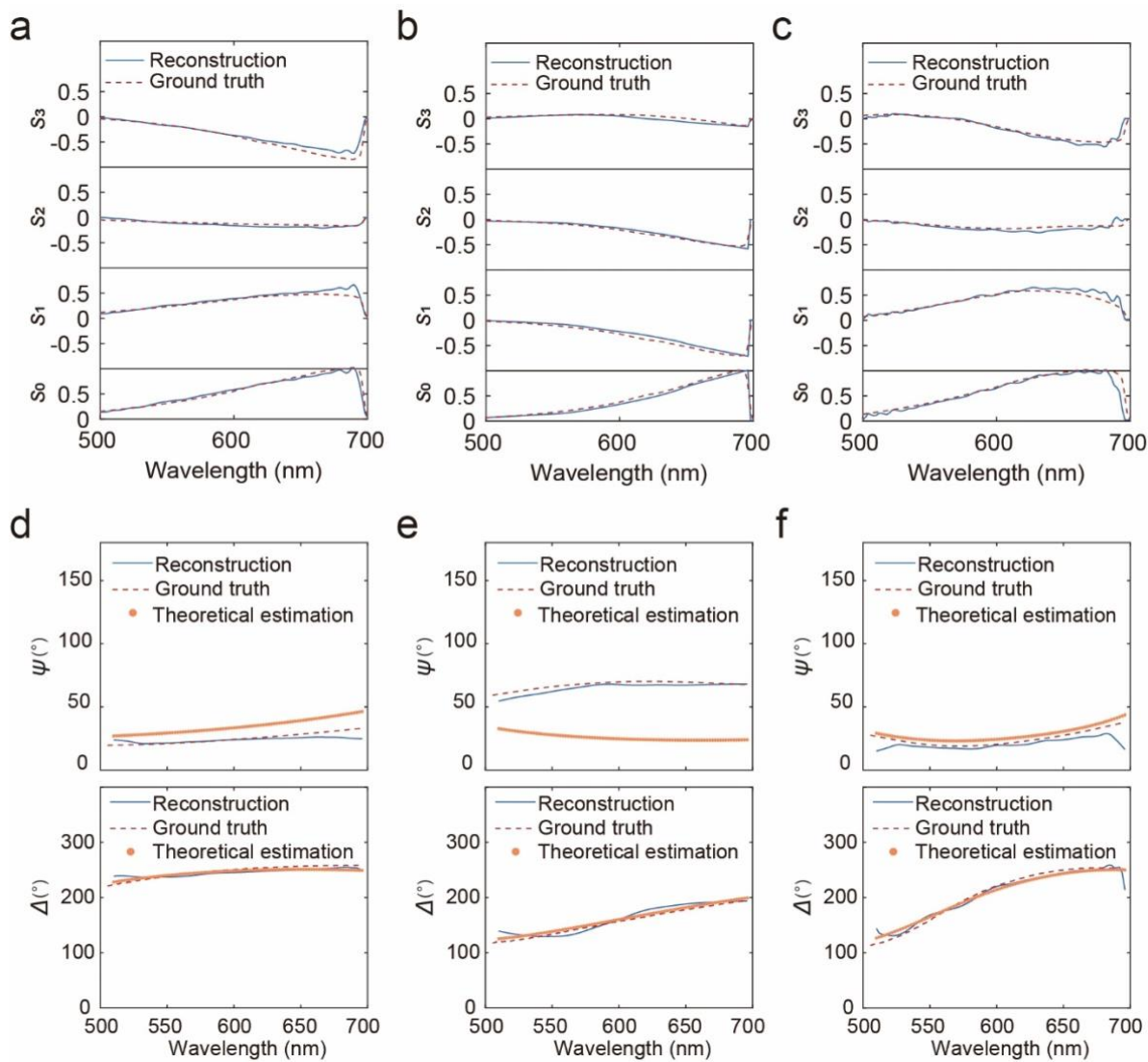

**Figure S10 | a-c,** Comparison between the reconstructed full Stokes polarization spectra from the single-shot spectroscopic ellipsometer (blue solid line), the ground truth (red dashed line) for the SiO$_2$ thin film with a thickness of 200 nm (a), 300 nm (b) and 500 nm (c), respectively. The ground truth is measured with a quarter-waveplate, a rotating polarizer, and a conventional grating-based spectrometer. **d-f**, Comparison among the ellipsometry parameters $\Psi$ and $\Delta$ from the single-shot spectroscopic ellipsometer (blue solid line), the ground truth (red dashed line) and the theoretical estimation (orange dots) for the SiO$_2$ thin film with a thickness of 200 nm (d), 300 nm (e) and 500 nm (f), respectively. The ground truth is measured with a quarter-waveplate, a rotating polarizer, and a conventional grating-based spectrometer.

## 11. The precision of the thickness and refractive index measurement for five SiO₂ thin films.

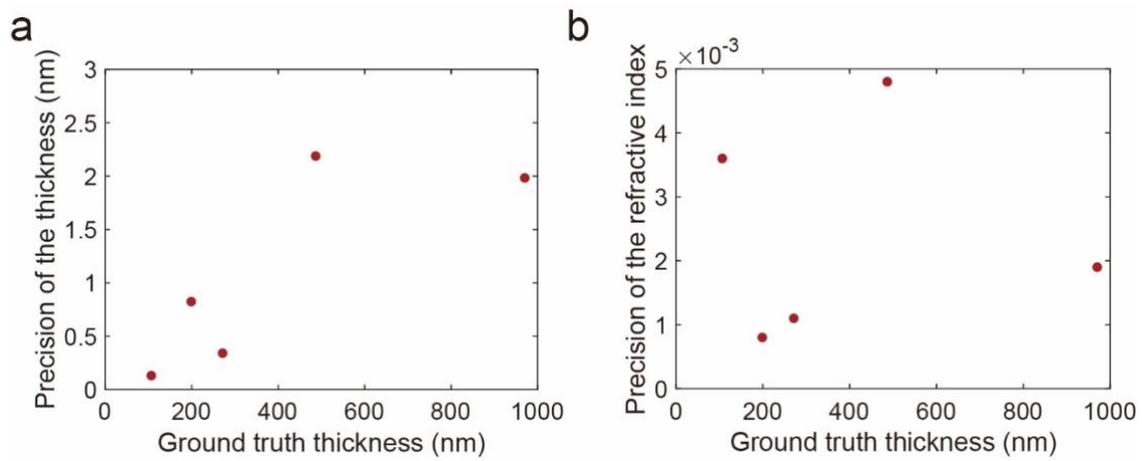

**Figure S11 | a**, Precision of the thickness measurement for five SiO$_2$ thin films. **b**, Precision of the refractive index measurement for five SiO$_2$ thin films.